\documentstyle{paper}
\begin{document}

\title{X-RAY MEASUREMENTS OF THE SHAPES OF CLUSTERS AND GALAXIES}

\author{Claude R. Canizares \& David A. Buote \\
{\it Department of Physics \& Center for Space Research,
Massachusetts Institute of Technology, Cambridge MA 02139, USA}}

\maketitle

\section*{Abstract}

We summarize a series of papers reporting studies of the ellipticities
of X-ray images of five Abell clusters and two early type galaxies to
learn about the shape of the gravitational potential and the
underlying matter distributions.  For the 
clusters we find the dark matter is centrally concentrated with ellipticity
$\epsilon_{DM} \sim 0.5$. For the galaxies we find independent
evidence for extended, flattened dark matter halos. One case of a
clear isophotal twist in the X-rays may indicate a triaxial halo.

\section{Introduction \& Method}

	The shapes of X-ray images of clusters and early type galaxies 
provide information about presence and distribution of dark matter in 
those systems (see Binney \& Strimpel 1978, Buote \& Canizares 1992, 
1994, 1996a, 1996b, 1996c, 1996d; Canizares \& White 1987).  Specifically, we 
can obtain independent, {\it geometrical evidence} for the presence of
dark matter  
(independent of the X-ray temperature, for example), giving a robust
test of the  
hypothesis that "mass follows light," and an independent measure of the 
ratio of luminous matter to dark matter.  We can also learn about the 
shape of the overall distribution of matter, probing the two-dimension 
flattening or three-dimensional triaxiality and placing independent 
constraints on the radial distribution.  Such information is valuable for 
constraining formation scenarios and for making comparisons to N-body 
codes, mass distributions deduced from gravitational lensing studies, 
and observations in other wavebands.

	The key assumption is that the X-ray emitting gas that we are 
observing in these systems is in or near hydrostatic equilibrium.  Recall 
that the timescales for equilibration in the centers of clusters and 
throughout elliptical galaxies are short compared to the lifetimes of the 
systems.  The major threat to the method for clusters is the possibility that 
equilibrium has been disturbed by a recent merger.
Other complications include the 
possible distortions in the outer parts of the interstellar medium (ISM) in 
galaxies due to external pressure from an intracluster medium (ICM), 
and possible emission from multiple phases in the ISM or ICM.  One can 
minimize these by choosing cases with relatively smooth images to avoid 
obvious recent mergers, searching for temperature anomalies and testing 
for reflection symmetry.  Also, since we use a low moment of 
the X-ray distribution, we are relatively insensitive to modest 
substructures.  For galaxies, the possibility of flattening due to rotation 
and contamination from point sources need to be considered explicitly.

	One can test susceptibility to substructure by exercising the 
method using N-body plus hydro simulations.  Buote \& Tsai (1995) 
analyzed a simulated cluster from Katz and White (1993) at various 
epochs.  At late epochs, $z\leq0.25$, when the simulated cluster looks 
relatively smooth, the X-ray gas appears to trace the gravitational 
potential, and our method does accurately measure the underlying 
matter distribution.  At earlier epochs the gas is not relaxed and the
X-rays
themselves trace the matter distribution rather than the potential.  

	As a reminder, the traditional application of hydrostatic 
equilibrium to analysis of X-ray images relies on the radial distributions 
of X-ray surface brightness and temperature to deduce the radial 
distribution of the total gravitating mass (e.g. see Bohringer, this 
volume).  The main difficulty is usually in obtaining a good temperature 
measurement at larger radii.

	Our method instead exploits the fact that for gas in hydrostatic 
equilibrium the isopotential surfaces are also the isodensity and 
isotemperature surfaces (BC94, BC96).  One can show that projection effects are 
generally not of concern, so that the projected X-ray surface brightness is 
a good tracer of the shape of the potential.  In particular this provides a 
robust test of the hypothesis that mass follows light .  

	Qualitatively, the isopotential surfaces of an elliptical spheroidal 
shell of matter (a so-called homoeoid) are themselves elliptical spheroids 
which are confocal to the homoeoid (Chandrasekar 1969, Binney \& 
Tremaine 1987).  Thus the ellipticity of the isopotentials falls rapidly with 
increasing distance from the homoeoid.  For example, the isopotentials of 
a mass shell with ellipticity $\epsilon\sim0.7$ and semi-major axis $a$ 
have $\epsilon\sim0.24$ at $1.5a$ and $\epsilon\sim0.06$ at $2.6a$ 
(see Binney \& Tremaine 1987, fig. 2-9).

	This means that if the visible light and X-rays come from roughly 
the same radius, as is roughly the case for clusters, then they should
have similar ellipticities; if not, one 
can conclude that the light cannot be tracing the mass.  Conversely, if 
the X-rays are more extended than the light, the X-ray isophotes must be 
nearly round if light traces mass; if the extended X-ray isophotes
show  sizeable 
ellipticity, this indicates presence of a dark matter component that is 
both elliptical and extended.

	Quantitatively, we construct an extensive grid of models based
on  a variety of 
assumed matter distributions, characterized by radial 
distribution and shape.  We compute the potential, fill it with gas and 
compare the resulting model X-ray distribution to the observations.
There  is only one adjustable  parameter 
beyond those that characterize the matter distribution, and that is the 
ratio of the specific gravitational energy to thermal energy at the center, 
$\Gamma={{\mu m\Phi_0}\over{kT_x}}$ , where $\mu m$ is the mean 
mass per particle, $\Phi_0$ is the central potential, $k$ is Boltzmann's 
constant and $T_x$ the gas temperature.  We can constrain $\Gamma$ 
observationally by fitting the observed X-ray radial profile; no
actual temperature measurement is required (although it can be used
separately with the derived $\Gamma$ to constrain $\Phi_0$). We iterate
many times to explore whole families of models and map out the 
permitted values of the shape parameters.  Generally
we assume the gas is isothermal, as is generally consistent with observations.
We have also explored temperature gradients, and find that they have
small effect on the shape of the permitted matter distributions but can alter
their radial profiles.

	One sidelight is that these shape studies provide interesting tests 
of the alternate explanation of the "dark matter problem" known as 
Modification of Newtonian Dynamics or MOND (e.g. Milgrom 1986).  MOND
attributes the well-know flat rotation curves to a break down of 
Newtonian dynamics at very small accelerations.  We have argued that 
MOND cannot explain mismatches in the {\it shapes} of visible and X-ray 
isophotes (BC94).

\section{Clusters of Galaxies}

	We applied our method to five, well known, rich Abell clusters, 
A401, A1656 (Coma), A2029, A2199, and A2256 (BC92, BC96b), all of 
which have flattened galaxy distributions on the sky, with 
$\epsilon \sim 0.4-0.6$.  We first used Einstein images (BC92), but the 
improved angular resolution of more recent ROSAT data has led us to 
modify our original conclusions.  The ROSAT data reveal a significant 
ellipticity gradient in four of the five clusters (the exception is Coma, 
which is large enough that we would not see a similar gradient).  
Examples are shown in Figure 1a.  Typically, we find 
$\epsilon_x\sim 0.25-0.3$ at $0.35h_{80}^{-1}$ Mpc falling to 
$\epsilon_x\sim 0.15$ at $1h_{80}^{-1}$ Mpc.  In all cases, the position 
angles of the X-ray and optical distributions are within the XXX degree 
uncertainties.

	One conclusion is that the matter giving rise to the
potential, which for clusters is
known to be mostly dark matter, is more centrally concentrated than the X-ray 
gas. Qualitatively, this follows from the fact that the X-rays and
galaxies occupying roughly the same regions have different
ellipticities.   The quantitative analysis supports this conclusion --
for example a  
distribution with density falling like $r^{-2}$ (as in a self-gravitating 
isothermal spheroid) cannot explain the data.  A flattened spheroidal 
mass distribution with $\epsilon\sim0.5$, as observed for the galaxies, but
falling more steeply, like $r^{-4}$, does work (the exact radial shape
depends on the assumed temperature gradient).  Therefore (contrary to our 
original conclusions in BC92 which assumed constant $\epsilon_x$) we 
find that the {\it shape} of the dark matter distribution can be similar to 
the shape of the visible matter, though they likely have different radial 
distributions.  Our analysis also gives revised values of gas and baryonic 
mass fractions that are consistent with previous results (e.g White \&
Fabian 1995).

\begin{figure}[t]
   \vspace{3.1 in}
   \caption{The observed ellipticity of X-ray isophotes for several 
Abell clusters vs. radius from the ROSAT PSPC (BC96b).} 
\end{figure}

\section{Early Type Galaxies}

	We studied two flattened, early type galaxies, inconclusively 
with Einstein (Canizares \& White 1987), then with ROSAT and now 
ASCA (BC94, BC95, BC96c). Optically, NGC 720 has $\epsilon\sim0.4$ and 
NGC 1332 has $\epsilon\sim0.7$.  Both are relatively isolated galaxies,
 and both are relatively slow
rotators.  Therefore they are unlikely to 
be significantly disturbed by external pressure or nearby companions.   Both
are at distances of approximately 20 Mpc. The best results are
obtained for NGC 720, which is brighter; NGC 1332 gives a consistent
picture at somewhat lower  significance.  

	We find X-ray ellipticities of $\epsilon_x \sim 0.2-0.3$ out to 
10 kpc and beyond.  As for most ellipticals, the visible light is very 
concentrated, with core radius $\leq 1 kpc$, so the potential from any mass 
distribution traced by the visible light would be nearly spherical at 10 
kpc.  Thus, the ellipticity of the X-ray isophotes alone constitutes 
{\it geometrical evidence} for the existence of a flattened dark matter halo.  
NGC720 shows an intriguing isophotal twist, described below, which is 
further evidence for dark matter.

	Quantitatively, we explored a wide range of models for comparison 
to the observations, adding progressively more and more dark halo until 
we can account for the observed X-ray ellipticities at large radii.  The 
models include a mass component distributed like the stars, a dark halo, 
and, in some cases, X-ray emission from discrete sources distributed like 
the stars.  We also considered effects of rotation at the stellar velocity, 
which are negligible for NGC 720 but could each account for approximately 
$\epsilon=0.1$ in NGC 1332.  We find that the dark halo must contain at 
least four times the mass component distributed like the stars (out to 
~25 kpc).  The overall mass to light ratio is ~30-50, and the ellipticity of 
the dark matter distribution is $\epsilon_{DM} \sim 0.5$.  

	ASCA is particularly helpful in constraining the possible 
contribution of discrete sources, which could be contributing to the 
apparent X-ray ellipticity in NGC 1332 (but is probably negligible in 
NGC 720).  Our observation shows clearly that NGC 1332 does have 
excess emission above ~3 keV that could not be from the hot, ISM and is 
most plausibly from the discrete source population (BC96d). This constitutes
approximately 25-50 \% of the 0.4-2.4 X-ray keV flux of the galaxy.

	Our original ROSAT PSPC image of NGC720 just barely resolved 
what appeared to be a "twist" of about 30 degrees in the isophotal 
position angle at a radius of about an arc minute (BC94).  We subsequently 
obtained ROSAT HRI data that confirms the twist (BC96c).  Figure 2 shows that 
the X-ray emission is aligned with the stars at small radii but then twists 
to the same value measured with the PSPC.  Dark matter, possibly in a 
triaxial distribution, is clearly indicated.

\begin{figure}[t]
   \vspace{3.1 in}
   \caption{The position angle of the elliptical isophotes of NGC 720
vs. radius from the ROSAT HRI (BC96d). The solid and dashed lines,
respectively, indicate the position angles and
approximate radial ranges for the
optical isophotes and for X-ray isophotes at larger radii from the
ROSAT PSPC.}   
\end{figure}

\section{Conclusion}

	Our analyses of the ellipticities of five flattened clusters and two 
flattened early type galaxies with ROSAT and ASCA lead us to the
following conclusions: 
\noindent For the clusters of galaxies:

\begin{itemize}
\item{the mass is not distributed like an isothermal sphere}
\item{the dark matter distribution is steep ($\sim r^{-4}$) at large
radii (for isothermal gas)}
\item{the dark matter is flattened with $\epsilon_{DM} \sim 0.5$}
\item{the ellipticity and position angle of the dark matter are consistent 
to those of the galaxy distribution}
\end{itemize}
\noindent For the early type galaxies:
\begin{itemize}
\item{the bulk of the mass does {\it not} follow the light}
\item{dark matter halo is extended and flattened}
\item{spectral studies constrain the contribution from discrete
sources to the observed X-ray ellipticity}
\item{the observed position angle twist may indicate triaxiality}
\end{itemize}

The cluster findings complement and are consistent with what is being
learned from 
gravitational lensing studies of arcs and arclets (e.g., Allen, Fabian
\& Kneib 1996, Pierre {\it et al.} 1996) and with expectations 
from N-body simulations (e.g., Efstathiou {\it et al.} 1988, Navarro,
Frenk \& White 1995). Similarly, the galaxy results add support to
we know about flattening from polar ring galaxies and agree with
expectations of  
dissipationless collapse (e.g., Dubinsky \& Carlberg 1991, Rix 1994,
Sacket 1995).  We believe 
that MOND cannot  
obviate this {\it geometrical evidence} for the presence of large
quantities of dark matter in clusters and galaxies.  
Finally, we suggest that the method can be powerfully applied to many 
more systems using the quality of data that will be obtained by AXAF.

\section{References}

\vspace{1pc}

\re
 	Allen, S., Fabian, A. \& Kneib, J.-P. 1996, {\it MNRAS} {\bf 279}, 615.
\re
 	Binney, J. \& Strimpel, O. 1978, {\it MNRAS} {\bf 187}, 473.
\re
 	Binney, J. \& Tremaine, S.  1987, {\it Galactic Dynamics}
(Princeton: Princeton Univ. Press)
\re
 	Buote, D. \& Canizares, C. 1992, {\it Ap. J.} {\bf 400}, 385 (BC92).
\re
 	Buote, D. \& Canizares, C. 1994, {\it Ap. J.} {\bf 427}, 86 (BC94).
\re
 	Buote, D. \& Canizares, C. 1996a, {\it Ap. J.} {\bf 457}, 177 (BC96a).
\re
 	Buote, D. \& Canizares, C. 1996b, {\it Ap. J.} {\bf 457}, 565 (BC96b).
\re
 	Buote, D. \& Canizares, C. 1996c, {\it Ap. J.} in press (BC96c).
\re
 	Buote, D. \& Canizares, C. 1996d, {\it Ap. J.} in press (BC96d).
\re
 	Buote, D. \& Tsai, J. 1995, {\it Ap. J.} {\bf 439}, 29.
\re
	Canizares, C. \& White, J. 1987 {\it BAAS} {\bf 19}, 682.
\re
	Chandrasekar, S. 1969  {\it Ellipsoidal Figures of
Equilibrium} (New Haven: Yale Univ. Press).
\re
 	Dubinski, J. \& Carlberg, R. 1991, {\it Ap. J.} {\bf 378}, 496.
\re
	Efstathious, G. {\it et al.} 1988  {\it MNRAS} {\bf 235}, 715.
\re
	Katz, N. \& White, S. 1993  {\it Ap. J.} {\bf 412}, 455.
\re
	Milgrom, M. 1986  {\it Ap. J.} {\bf 302}, 617.
\re
	Navarro, J. Frenk, C. \& White, S. 1995  {\it A\&A} {\bf 275} 720.
\re
	Pierre, M. {\it et al.} 1996  {\it A\&A} in press.
\re
	Rix, H-W. 1994 in {\it IAU Symp. 169 Unsolved Problems of the
Milky Way}.
\re
	Sackett, P. 1995 in Kochaneck, C. \& Hewitt, J. (eds), {\it
IAU Symp. 173 Gravitational Lensing}.
\re
 	White, D. \& Fabian, A. 1995, {\it MNRAS} {\bf 273}, 72.

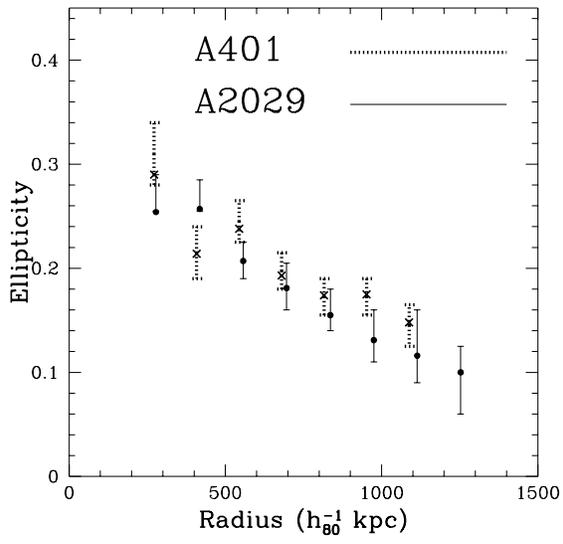
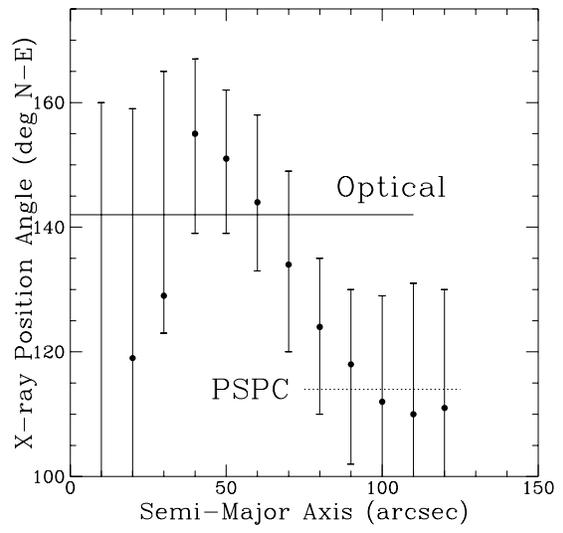
\end{document}